\documentclass[%
twocolumn,
amsmath,amssymb,
aps,
superscriptaddress,
]{revtex4-2}

\usepackage{url}
\usepackage[dvipdfmx]{graphicx}
\usepackage{afterpage}
\usepackage{siunitx}
\usepackage{amsmath, amssymb}
\usepackage{natbib}
\DeclareSIUnit\angstrom{\text {Å}}
\usepackage{dcolumn}
\usepackage{bm}
\usepackage{tikz}
\usetikzlibrary{positioning,intersections,angles,quotes}
\usepackage{booktabs} 
\usepackage{comment}
\usepackage{physics}
\usepackage{mathtools}
\usepackage{braket}
\usepackage{algorithm}
\usepackage{algpseudocode}
\usepackage{float}

\usepackage[colorlinks=true,citecolor=blue,linkcolor=magenta]{hyperref}

\usepackage{longtable}
\usepackage{booktabs}
\usepackage{svg}

\begin{document}

\title{QAOA-MC: Markov chain Monte Carlo enhanced by \\ Quantum Alternating Operator Ansatz}

\author{Yuchiro Nakano}
\email{u830977g@ecs.osaka-u.ac.jp}
\affiliation{
    Graduate School of Engineering Science, Osaka University, 1-3 Machikaneyama, Toyonaka, Osaka 560-8531, Japan.
}

\author{Hideaki Hakoshima}
\email{hakoshima.hideaki.qiqb@osaka-u.ac.jp}
\affiliation{
    Center for Quantum Information and Quantum Biology, Osaka University, 560-0043, Japan.
}

\author{Kosuke Mitarai}
\email{mitarai.kosuke.es@osaka-u.ac.jp}
\affiliation{
    Graduate School of Engineering Science, Osaka University, 1-3 Machikaneyama, Toyonaka, Osaka 560-8531, Japan.
}
\affiliation{
    Center for Quantum Information and Quantum Biology, Osaka University, 560-0043, Japan.
}

\author{Keisuke Fujii}
\email{fujii.keisuke.es@osaka-u.ac.jp}
\affiliation{
    Graduate School of Engineering Science, Osaka University, 1-3 Machikaneyama, Toyonaka, Osaka 560-8531, Japan.
}
\affiliation{
    Center for Quantum Information and Quantum Biology, Osaka University, 560-0043, Japan.
}
\affiliation{
    Center for Quantum Computing, RIKEN, Wako Saitama 351-0198, Japan.
}

\date{\today}

\begin{abstract}
    Quantum computation is expected to accelerate certain computational task over classical counterpart. Its most primitive advantage is its ability to sample from classically intractable probability distributions. A promising approach to make use of this fact is the so-called quantum-enhanced Markov chain Monte Carlo (MCMC) [D. Layden, \textit{et al.}, arXiv:2203.12497 (2022)] which uses outputs from quantum circuits as the proposal distributions. 
    In this work, we propose the use of Quantum Alternating Operator Ansatz (QAOA) for quantum-enhanced MCMC
    and provide a strategy to optimize its parameter to improve convergence speed
    while keeping its depth shallow. 
    The proposed QAOA-type circuit is designed to satisfy the specific constraint which quantum-enhanced MCMC requires with arbitrary parameters. 
    Through our extensive numerical analysis, we find a correlation in certain parameter range between an experimentally measurable value,  acceptance rate of MCMC, and the spectral gap of the MCMC transition matrix, which determines the convergence speed.
    This allows us to optimize the parameter in the QAOA circuit and achieve quadratic speedup in convergence.
    Since MCMC is used in various areas such as statistical physics and machine learning makes, 
    this work represents an important step toward realizing practical quantum advantage with currently available quantum computers through quantum-enhanced MCMC.
\end{abstract}

\maketitle

\section{Introduction}
The number of qubits in current quantum computers is limited.
They hence lack the capacity to implement quantum error correction, rendering them vulnerable to noise.
These emerging quantum computing systems are called Noisy Intermediate-Scale Quantum (NISQ) devices \cite{Preskill_2018}. 
These devices have successfully demonstrated the superiority of quantum computers over classical computers in practice: ``quantum supremacy'' \cite{Arute_2019,madsen2022quantum}.
For practical applications, Variational Quantum Algorithms (VQAs) \cite{Cerezo_2021} emerge as a promising approach to utilize the NISQ devices.
VQAs run the parameterized quantum circuits (called variational quantum circuits) on NISQ devices and optimize the parameter using an objective function which is expressed by an expected value of an observable with respect to the output distribution.
This approach keeps the quantum circuit depth shallow because the optimization is performed on classical computers.
Some algorithms based on the VQA framework have been proposed for quantum chemical computation \cite{Peruzzo_2014}, combinatorial optimization \cite{Farhi_2014}, and machine learning \cite{Mitarai_2018, farhi2018classification}.

Unfortunately, existing VQAs have yet to demonstrate a quantum advantage over the state-of-the-art classical approach for solving those problems.
A possible weakness of these algorithms is their use of the expected values of operators.
This requires us to run quantum circuits many times to suppress statistical errors.
Furthermore, since optimization using the expected value is performed iteratively, the total runtime can be prohibitively large \cite{K_bler_2020,Gonthier_2022,ito2023latencyaware}.

In contrast, algorithms that utilize each sampling outputs from quantum circuits may be more suitable for making use of NISQ devices.
For example, Random Circuit Sampling (RCS) used to demonstrate quantum supremacy in NISQ devices is the task of sampling from the output distribution of a random quantum circuit\cite{hangleiter2023computational} and is shown to be classically hard under a plausible conjecture \cite{aaronson2016complexitytheoretic}.
Sampling from Instantaneous Quantum Polynomial (IQP) circuits \cite{Bremner2017achievingquantum} and random linear optical circuits \cite{aaronson2011computational} are other famous examples whose classical hardness is strongly believed.
These examples motivate us to develop algorithms that fully exploit each sampling outcome from NISQ devices.

The quantum-enhanced Markov chain Monte Carlo (quantum-enhanced MCMC) method \cite{Layden_2022} is one of such algorithms, which uses samples from a quantum circuit as the proposal distribution in the Metropolis-Hastings method \cite{Hastings_1970}.
Markov chain Monte Carlo (MCMC) \cite{Metropolis_1953,Hastings_1970} is a powerful technique for sampling from computationally difficult distributions such as the Boltzmann distribution and has many applications in statistical physics \cite{landau_binder_2014}, combinatorial optimization \cite{kirkpatrick1983optimization}, and machine learning \cite{andrieu2003introduction}.
The Metropolis-Hastings method, one of the MCMC methods, consists of two steps: the generation of a sample by the proposal distribution and the accepting or rejecting step of this sample.
Since the proposal distribution determines the efficiency of the algorithm, the proposal distribution using a quantum computer, including those that are difficult to simulate on classical computers, can improve the convergence speed of MCMC over existing ones.
The quantum-enhanced MCMC uses distribution defined by a classically intractable quantum state as the proposal.

The circuit proposed in Ref.~\cite{Layden_2022} is expressed as the time evolution governed by a time-independent Hamiltonian.
However, when running it on quantum computers, the time evolution has to be decomposed by the Suzuki-Trotter expansion \cite{suzuki1976generalized}, which increases the circuit depth when the evolution time is long.
This fact makes implementation of the quantum-enhanced MCMC on NISQ devices infeasible depending on the choice of the time parameter.
In addition, the quantum circuits and their parameters are selected heuristically, and the strategy to construct a quantum circuit that improves the convergence speed of MCMC remains unclear.

In this paper, based on the quantum-enhanced MCMC, we propose a new MCMC method called Quantum Alternating Operator Ansatz Monte Carlo (QAOA-MC).
This algorithm uses a fixed-depth variational quantum circuit in the form of the so-called Quantum Alternating Operator Ansatz (QAOA) \cite{hadfield2019quantum} as the proposal distribution.
We thereby aim to suppress the increase in circuit depth regardless of the choice of parameters.
Furthermore, we construct a systematic strategy to optimize the circuit to improve the convergence speed by examining the relationship between the absolute spectral gap and the acceptance rate (AR) of the proposal distribution. 
More precisely, we find that MCMC can be accelerated by minimizing AR after properly limiting the parameter range and reducing the number of circuit parameters.
Through the numerical experiments, we evaluate the performance of our method through the Boltzmann distribution for a spin glass model, which is one of the most challenging systems to simulate due to its complex energy landscape and slow relaxation dynamics. 
As a result, we show that QAOA-MC achieves a near quadratic speed-up in the convergence speed compared to the proposal using the uniform distribution.
Additionally, we demonstrate QAOA-MC through the estimation of the average magnetization in a spin glass consisting of 15 spins.
Our results suggest an acceleration of MCMC using NISQ devices and contribute to promoting the use of current NISQ devices.

The rest of the paper is organized as follows.
First, we will explain MCMC in detail and introduce the quantum-enhanced MCMC in Sec.~\ref{preliminary}.
Furthermore, we discuss the challenges of the quantum-enhanced MCMC.
Sec.~\ref{Method} outlines our scheme: QAOA-MC.
There, we propose to utilize a variational quantum circuit for MCMC proposals and provide guidance on optimizing the circuit. 
In Sec.~\ref{Numerical_Experiments}, we describe the details of the numerical experiments and their results.
Finally, a conclusion and future perspectives are presented in Sec.~\ref{Conclusion}.

\section{Preliminary}
\label{preliminary}
In this section, we provide an overview of MCMC and introduce the quantum-enhanced MCMC.

\subsection{Markov chain Monte Carlo (MCMC)}
\label{Markov_Chain_Monte_Carlo}
Markov chain Monte Carlo (MCMC) method is a very powerful algorithm that can sample according to an arbitrary probability distribution.
This algorithm starts with a state $\bm{x} = [ x_1, x_2, \cdots, x_n ]$ and changes the state according to a Markov chain, which is a stochastic process denoted by a transition probability $P(\bm{x}'|\bm{x})$.
For this Markov chain to converge to the desired a target distribution $\pi(\bm{x})$, it must be irreducible and aperiodic \cite{Levin_2017}.
The most general way to satisfy these conditions is by imposing the detailed balance to $P(\bm{x}'|\bm{x})$. 
The detailed balance is that for any state transition from $\bm{x}$ to $\bm{x}'$, the following equation is satisfied:
\begin{align}
    \pi(\bm{x}') P(\bm{x}' | \bm{x}) = \pi(\bm{x}) P(\bm{x} | \bm{x}') \quad \forall \bm{x}, \bm{x}' .
    \label{eq:DetailedBalance}
\end{align}
The Metropolis-Hastings method \cite{Metropolis_1953,Hastings_1970} realizes a transition that satisfies detailed balance.
In this method, a transition from $\bm{x}$ to $\bm{x}'$ with probability $P(\bm{x}'|\bm{x})$ is factored into the proposal distribution $Q(\bm{x}'|\bm{x})$ and the acceptance probability $A(\bm{x}'|\bm{x})$ for the proposal.
The procedure is as follows: first, propose the next state according to $Q(\bm{x}'|\bm{x})$.
Secondly, decide whether to accept the proposal based on the acceptance probability \begin{align}
    A(\bm{x}'|\bm{x}) = \min \left( 1, \frac{\pi(\bm{x}')}{\pi(\bm{x})} \frac{Q(\bm{x}|\bm{x}')}{Q(\bm{x}'|\bm{x})} \right).
    \label{eq:MH_acceptance}
\end{align}
If the proposal is rejected, the next state remains the same as the state prior to the proposal.
There are no restrictions on the choice of $Q(\bm{x}'|\bm{x})$, but $Q(\bm{x}'|\bm{x})$ must be in a form that can be calculated efficiently.

In MCMC, the choice of proposal distributions is crucial as it determines the convergence speed.
Here, we briefly review some proposal strategies in the case of spin systems.
The simplest method is to flip one spin in the configuration at random, which is called the local update.
This method can be applied to any model and is easy to implement.
However, it can increase the computational time depending on the shape of $\pi(\bm{x})$.
This problem can be avoided by flipping multiple spins at once. 
This proposal is called the global update.
One of the simplest global updates is to propose transitions with equal probability for all possible states. 
This proposal follows a uniform distribution, which we refer to as the ``uniform update''.
A more sophisticated global update is called the cluster update. 
In this update, we flip all spins in a group (a cluster), which is determined according to model-specific algorithms.
The cluster update improves computational time for certain models \cite{Swendsen_1987,Wolff_1989}. 
However, generating clusters using this method is not straightforward and it can only be applied to specific models.

Finally, we will now describe the convergence speed of MCMC.
The convergence speed of a Markov chain can be represented by the eigenvalues of the transition matrix $P$ \cite{Peskun_1973,frigessi1992optimal}.
In this work, we use a quantity called absolute spectral gap as the metric to evaluate convergence speed following Ref.~\cite{Layden_2022}.
The absolute spectral gap \cite{Levin_2017} is defined as the absolute difference between the first two largest eigenvalues ($\lambda_1$ and $\lambda_2$) of the transition probability matrix $P$ of a Markov chain, which is represented by
\begin{align}\label{eq:spectral-gap}
    \delta = 1 - |\lambda_2|.
\end{align}
$\delta$ is in the range from 0 to 1.
The larger the value of $\delta$, the faster the convergence speed becomes. 

\subsection{Quantum-enhanced MCMC}
\label{Quantum_Algorithm}
The quantum-enhanced MCMC algorithm, developed by Layden et al. \cite{Layden_2022}, samples the Boltzmann distribution for the classical Ising model using NISQ devices.
The Boltzmann distribution describes the thermal equilibrium state of a system at temperature $T$ and is defined as
\begin{align}
    \mu(\bm{x}) &= \frac{1}{Z} \exp \left( -\frac{E(\bm{x})}{T} \right), \\
    Z &= \sum_{\bm{x}} \exp \left( -\frac{E({\bm{x}})}{T} \right).
    \label{eq:Boltzmann_dist}
\end{align}
Here, $Z$ represents a partition function: the sum of the Boltzmann factor $\exp(-E({\bm{x}}_j)/T)$ for all states ${\bm{x}}$ and $E(\bm{x})$ is the energy of the state $\bm{x}$.

The quantum-enhanced MCMC algorithm uses a quantum circuit to sample a proposal distribution to realize the probability distribution $\mu(\bm{x})$.
More concretely, the proposal is executed by applying the quantum circuit $U$ to $\ket{\bm{x}}$ which encodes $\bm{x}$ as a quantum state, and then measuring $U\ket{\bm{x}}$ to obtain $\bm{x}'$.
The proposal distribution is therefore given by $Q(\bm{x}'|\bm{x}) = {|\braket{\bm{x}' | U | \bm{x}} |}^2$.
At first sight, Eq.~\eqref{eq:MH_acceptance} seems to require us to compute the exact value of $Q(\bm{x}'|\bm{x})$.
The computation of ${|\braket{\bm{x}' | U | \bm{x}} |}^2$ for a general quantum circuit $U$ generally requires exponential time and cannot be performed efficiently even with quantum computers.
However, we can avoid its computation by imposing the symmetry of $U=U^{\top}$ on the quantum circuit, which leads to:
\begin{align}
    Q(\bm{x}'|\bm{x}) = {| \braket{\bm{x}' | U | \bm{x}} |}^2 = {| \braket{\bm{x} | U | \bm{x}'} |}^2 = Q(\bm{x}|\bm{x}').
\end{align}
This cancels out the $Q$ term in Eq.~(\ref{eq:MH_acceptance}) and reduces to a simpler method called the Metropolis method \cite{Metropolis_1953}. 
When the target distribution is the Boltzmann distribution $\mu(\bm{x})$, Eq.~(\ref{eq:MH_acceptance}) becomes:
\begin{align}
    A(\bm{x}'|\bm{x}) = \min \left[ 1, \exp \left( \frac{E(\bm{x})-E(\bm{x}')}{T} \right) \right],
    \label{Metropolis}
\end{align}
which can be efficiently calculated on a classical computer.
The quantum circuit is only used for the proposal $\bm{x} \rightarrow \bm{x }'$.
The calculation of $A(\bm{x}'|\bm{x})$ as well as the decision of accepting/rejecting the proposal is done on classical computers.
More importantly, while VQAs necessitate multiple runs and measurements of the quantum circuit to calculate the objective function and optimize the circuit, the quantum-enhanced MCMC requires only a single run and measurement of the quantum circuit for each MCMC step.

A variety of quantum circuits can be used in the quantum-enhanced MCMC algorithm as long as $U=U^{\top}$ is satisfied.
In Ref.~\cite{Layden_2022}, they have used the time evolution under a time-independent Hamiltonian $H$ as $U$ and evaluated the performance of this algorithm.
More concretely, their choice of $U$ is given by
\begin{align}
    \label{eq:qe_mcmc_u}
    U &= \exp(-iHt), \\
    H &= (1-u) \alpha H_{\rm{prob}} + u H_{\rm{mix}},
\end{align}
where 
\begin{align}
    \label{eq:alpha}
    \alpha = {|| H_{\rm{mix}} ||}_{\rm{F}} / {|| H_{\rm{prob}} ||}_{\rm{F}}
\end{align}
is the normalization factor for $H_{\rm{mix}}$ and $u \in [0,1]$ is a parameter that controls the relative weights of both $H_{\rm{mix}}$ and $H_{\rm{prob}}$.
The $H_{\rm{mix}}$ and $H_{\rm{prob}}$ are given by
\begin{align}
    H_{\rm{mix}} &= \sum_{j=1}^n X_j,
    \label{eq:mixer_hamiltonian}
    \\
    H_{\rm{prob}} &= - \sum_{\braket{j,k}} J_{jk} Z_j Z_k - \sum_{j=1}^n h_j Z_j,
    \label{eq:problem_hamiltonian}
\end{align}
where $X_j$ and $Z_j$ are the Pauli operators acting on the $j$-th qubit, and $H_{\rm prob}$ is the target Hamiltonian from whose Boltzmann distribution we wish to sample.
The coefficients $\{ J_{jk} \}$ and $\{ h_j \}$ are defined by couplings and external fields of the target Hamiltonian.
The algorithm flow is shown in Algorithm~\ref{qe_mcmc_algorithm}.
\begin{algorithm}[H]
    \caption{Quantum-enhanced MCMC \cite{Layden_2022}}
    \label{qe_mcmc_algorithm}
    \begin{algorithmic}[1]
        \State $\bm{x} = \text{initial spin configuration}$
        \While {$\text{not converged}$} 
            \State $\textbf{Propose jump (quantum step)}$
            \State $u = \text{random.uniform(0.25, 0.6)}$
            \State $t = \text{random.uniform(2, 20)}$
            \State $\ket{\psi} = \exp[-iH(u)t]\ket{\bm{x}} \text{on quantum device}$
            \State $\bm{x'} = \text{result of measuring} \ket{\psi} \text{in computational basis}$
            \Statex
            \State $\textbf{Accept/reject jump (classical step)}$
            \State $A = \min(1, \exp[(E(\bm{x})-E(\bm{x'}))/T] )$
            \If {$A \geq \text{random.uniform(0, 1)}$}
                \State $\bm{x} = \bm{x'}$
            \EndIf
        \EndWhile
    \end{algorithmic}
\end{algorithm}

However, $U$ in this form must be implemented by the Suzuki-Trotter decomposition, which increases the circuit depth depending on the choice of the time parameter $t$. 
In addition, the parameters $(u,t)$ are chosen randomly, and no optimization method has been established.
This work aims to resolve these challenges.

\section{MCMC with Variationally Trained Quantum Sampling}
\label{Method}
In this section, based on the quantum-enhanced MCMC, we propose a new MCMC method that introduces a variational quantum circuit as the proposal distribution and optimizes this parameter by using the MCMC acceptance rate. We call this method Quantum Alternating Operator Ansatz Monte Carlo (QAOA-MC).

\subsection{Variational quantum circuit}
\begin{figure*}[t]
    \centering
    \includegraphics[width=\linewidth]{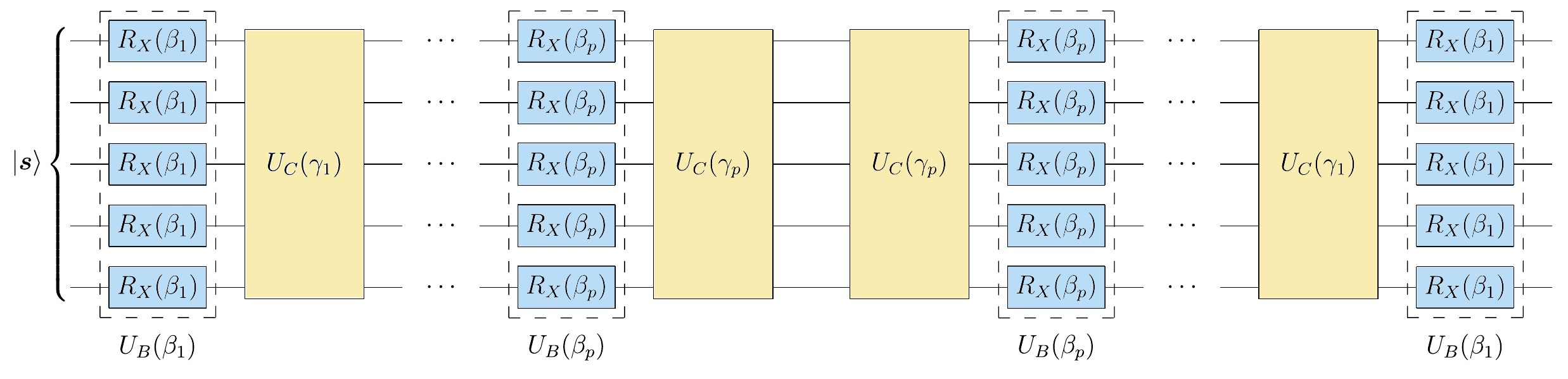}
    \caption{The variational quantum circuit used in our algorithm.}
    \label{quantum_circuit}
\end{figure*}
We apply the Quantum Alternating Operator Ansatz (QAOA) \cite{hadfield2019quantum} to the structure of the circuit that generates MCMC proposals.
Concretely, our circuit is defined as follows:
\begin{align}
    U = {V(\bm{\beta}, \bm{\gamma})}^\top V(\bm{\beta}, \bm{\gamma}),
    \label{eq:u_proposed}
\end{align}
where,
\begin{align}
    V(\bm{\beta}, \bm{\gamma}) = U_C(\gamma_p)U_B(\beta_p) \cdots U_C(\gamma_1)U_B(\beta_1), \\
    U_B(\beta) = \exp(-i H_{\rm{mix}} \beta) , \quad U_C(\gamma) = \exp(-i \alpha H_{\rm{prob}} \gamma),
\end{align}
and $p$ is a hyperparameter that determines the depth of the circuit.
It is shown in Fig.~\ref{quantum_circuit}.
The circuit has $2p$ parameters: $\bm{\beta} = \{ \beta_1, \cdots, \beta_p \}$ and $\bm{\gamma} = \{ \gamma_1, \cdots, \gamma_p \}$. 
$\alpha$, $H_{\rm{mix}}$, and  $H_{\rm{prob}}$ are as defined in Eqs.~\eqref{eq:alpha}, \eqref{eq:mixer_hamiltonian}, and \eqref{eq:problem_hamiltonian}. 
Note that the circuit defined above always satisfies $U=U^{\top}$ by construction.

This circuit has a similar structure to Eq.~\eqref{eq:qe_mcmc_u}.
However, unlike the circuit implementation in Eq.~\eqref{eq:qe_mcmc_u}, our circuit is more NISQ-friendly because the circuit depth is fixed.
The initial state $\ket{\bm{x}}$ is expected to be updated globally through $H_{\rm{mix}}$ and $H_{\rm prob}$ is responsible for proposing a transition respecting the energy landscape of the system.
The generated probability distribution $Q(\bm{x}'|\bm{x})$ includes those that are classically difficult to simulate and may realize acceleration of the convergence compared to existing proposal distributions.

\subsection{Optimization of circuit}
Next, we explain how to optimize the proposal distribution generated by the proposed circuit (Eq.~\eqref{eq:u_proposed}) to achieve faster convergence.
One might think that we can use the absolute spectral gap $\delta$ (Eq.~\eqref{eq:spectral-gap}) which directly determines the convergence speed as an objective function to maximize.
However, computing $\delta$ requires solving for the eigenvalues of a transition matrix with a size of $2^n \times 2^n$ for a system of size $n$ and is not feasible.
The objective function must be a quantity that reflects the convergence speed of MCMC and is easily computable.
We find that the MCMC acceptance rate can be used as the objective function after some numerical experiments.

The acceptance rate (AR) \cite{Roberts_1997} is defined as:
\begin{align}
    {\rm{AR}} = \sum_{\bm{x},\bm{x}'} \pi({\bm{x}}) Q({\bm{x}}'|{\bm{x}}) A({\bm{x}}'|{\bm{x}}).
    \label{eq:AR}
\end{align}
This formula includes $\pi(\bm{x})$, making it difficult to be calculated directly.
However, it can efficiently be estimated by performing MCMC on $\pi(\bm{x})$.
We estimate $\mathrm{AR}$ using samples generated by $M$ MCMC steps as follows:
\begin{align}
    \label{eq:MCMC-AR}
    {\rm{AR}} \approx \frac{1}{M} \sum_{j=0}^{M-1} A({\bm{x}}^{(j+1)}|{\bm{x}}^{(j)}),
\end{align}
where ${\bm{x}}^{(j)}$ represents the state at the $j$-th step of the MCMC.

\begin{figure}
    \centering
    \includegraphics[width=\linewidth]{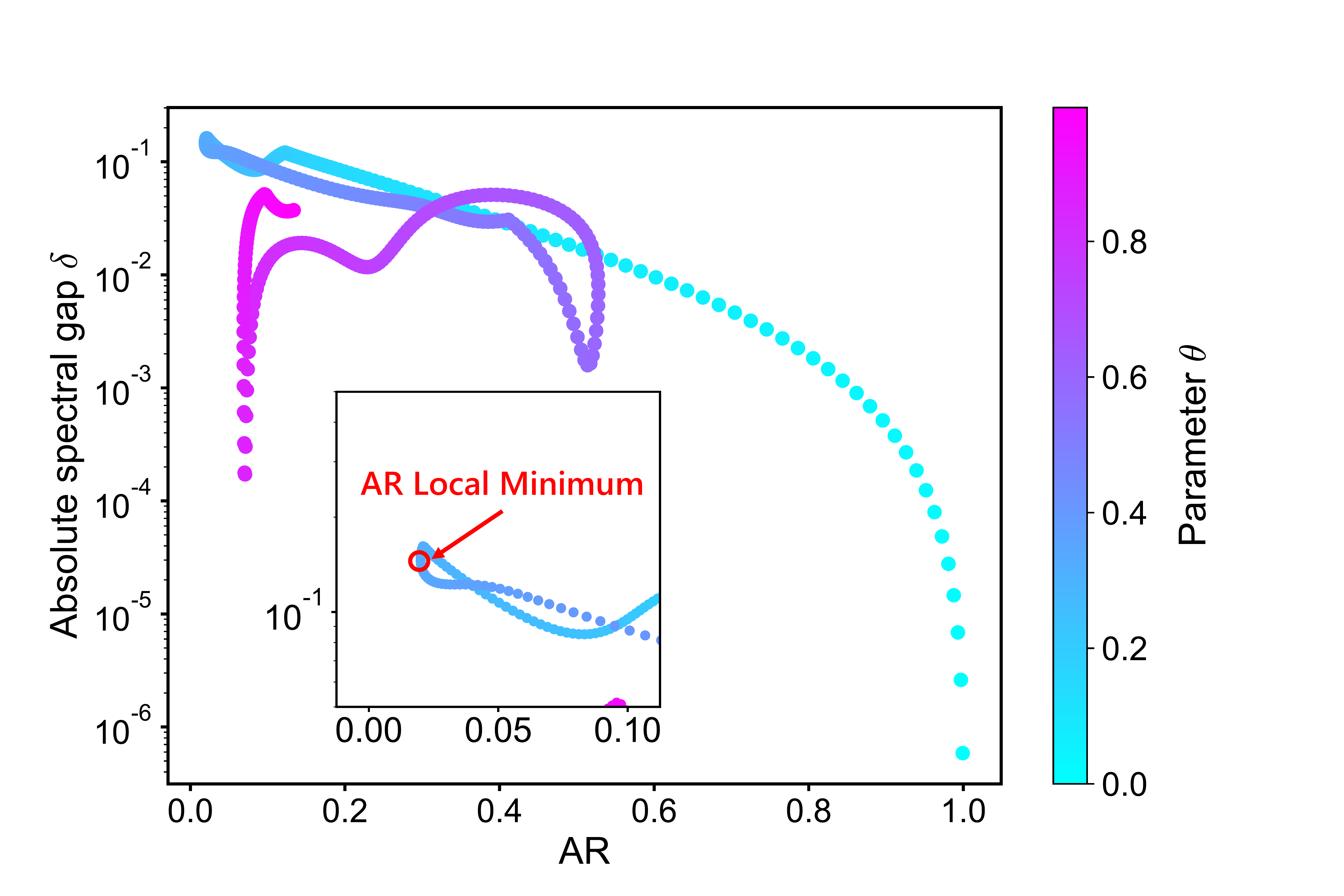}
    \caption{The relationship between AR and the absolute spectral gap $\delta$ in a typical instance.
    In many cases, the smallest parameter among those taking AR minima gives a large absolute spectral gap.}
    \label{typical_parameter_varies}
\end{figure}

After experimenting with the Boltzmann distributions for various Ising models, we have discovered a relationship between AR and the absolute spectral gap $\delta$.
Figure~\ref{typical_parameter_varies} illustrates the relationship between $\delta$ and AR in a typical Ising model instance.
In this experiment, we use a single-parameter circuit $U(\theta)$ which is defined by setting the parameters in Eq.~\eqref{eq:u_proposed} as
\begin{align}
    \theta = \beta_1 = \cdots = \beta_p = \gamma_1 = \cdots = \gamma_p . \notag
\end{align}
Although there is no correlation between AR and $\delta$ in general, a correlation exists for small $\theta$, where $\delta$ increases as AR decreases.
This continues until the AR reaches a local minimum, which often produces a local maximum value of $\delta$.
Based on these observations, we optimize $U(\theta)$ by searching for a small $\theta$ that achieves the locally minimal AR.

This observation may seem counterintuitive.
Since rejecting as few proposals as possible is intuitively preferable for achieving faster convergence, maximizing, rather than minimizing, AR may be considered as the strategy to find the proposal distribution.
However, as noted in Ref.~\cite{Neklyudov_2018}, for the Metropolis-Hastings method, maximizing AR does not necessarily contribute to speeding up the convergence because the ``delta-function'' proposal which proposes the same state as before with probability 1 also has $\rm{AR}=1$.

Finally, our algorithm flow is shown in Algorithm~\ref{proposed_algorithm}.
We perform a search for the local minimum value of AR by restricting the search range to $\theta \in (0, \theta_{\rm{max}}]$.
$\theta_{\rm{max}}$ is set as a hyperparameter in our algorithm.
We find that a fixed $\theta_{\rm max}$ can be used for different model instances without deteriorating the performance as long as we fix the depth parameter $p$ (see Appendix~\ref{appendix}).
\begin{algorithm}[H]
    \caption{QAOA-MC}
    \label{proposed_algorithm}
    \begin{algorithmic}[1]
        \State $\textbf{Optimization of parameter by minimizing AR}$
        \State $\text{Set} \: \theta = 0.01$
        \While {$\theta \: \text{not converged}$} 
            \State $\text{Estimate AR at $\theta$}$
            \State $\text{Optimize $\theta$ using AR estimator}$
        \EndWhile
        \State $\text{Get an optimized parameter $\theta^*$}$
        \Statex
        \State $\textbf{Main MCMC simulation}$
        \State $\bm{x} = \text{initial spin configuration}$
        \While {$\text{not converged}$} 
            \State $\textbf{Propose jump (quantum step)}$
            \State $\ket{\psi} = U(\theta^*)\ket{\bm{x}} \text{on quantum device}$
            \State $\bm{x'} = \text{result of measuring} \ket{\psi} \text{in computational basis}$
            \Statex
            \State $\textbf{Accept/reject jump (classical step)}$
            \State $A = \min(1, \exp[(E(\bm{x})-E(\bm{x'}))/T] )$
            \If {$A \geq \text{random.uniform(0, 1)}$}
                \State $\bm{x} = \bm{x'}$
            \EndIf
        \EndWhile
    \end{algorithmic}
\end{algorithm}

\section{Numerical Experiments}
\label{Numerical_Experiments}

\subsection{Model}
\label{model}
\begin{figure}
    \centering
    \includegraphics[width=\linewidth]{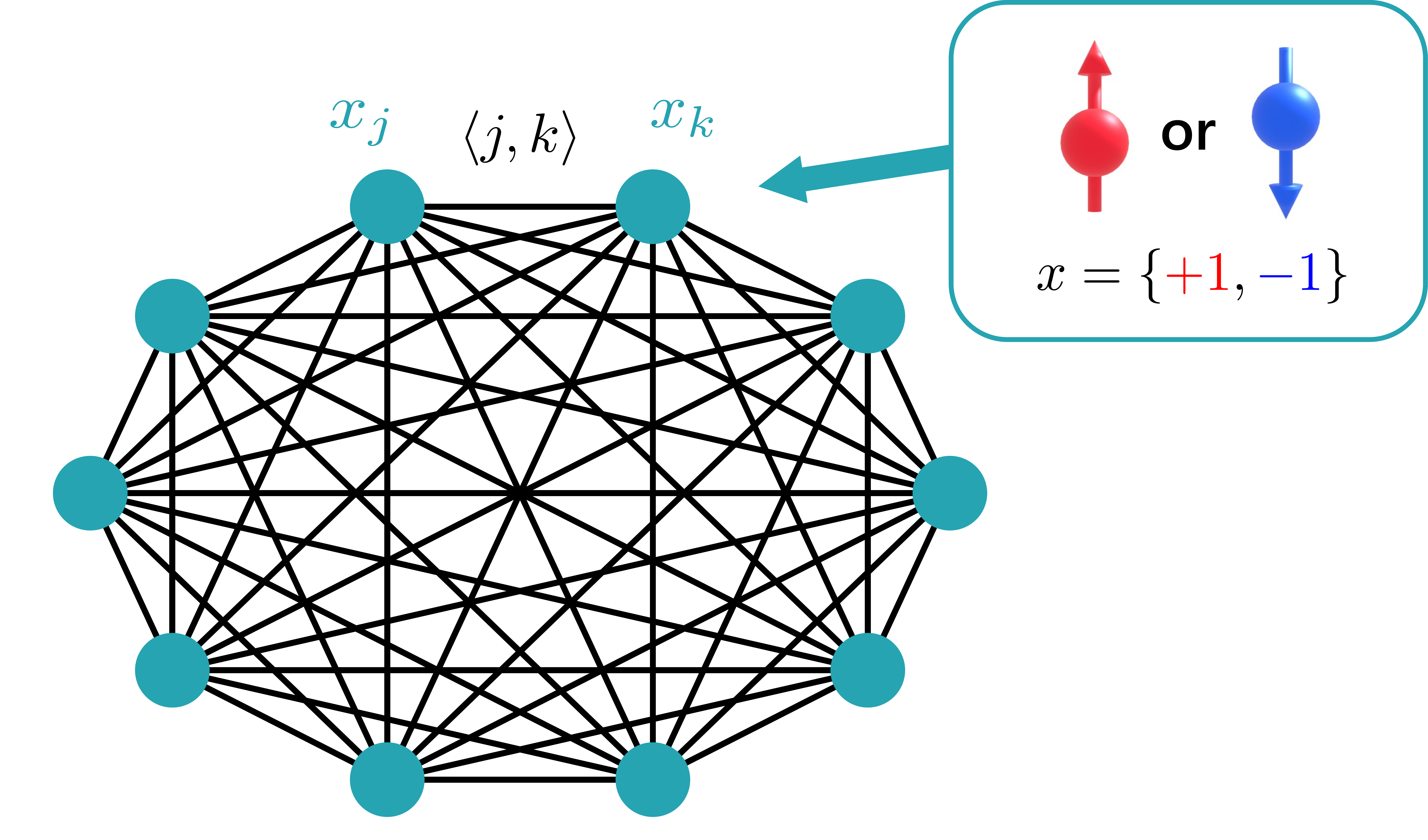}
    \caption{An $n=10$ example of the spin glass instance (Eq.~\eqref{eq:ising}). The spins $x_j$ and $x_k$ at each site take a value of either $+1$ or $-1$ depending on the upward and downward direction. The couplings between $x_j$ and $x_k$ are represented by the edges $\braket{j,k}$ in the graph. This model has all-to-all coupling.}
    \label{ising_instance}
\end{figure}

In all our numerical experiments, we use the Boltzmann distribution $\mu(\bm{x})$ (Eq.~\eqref{eq:Boltzmann_dist}) for a spin glass for the target distribution.
The energy function of the spin glass is
\begin{align}
    E(\bm{x}) = -\sum_{j>k=1}^n J_{jk} x_j x_k - \sum_{j=1}^n h_j x_j,
    \label{eq:ising}
\end{align}
where $x_j \in \{1, -1\}$ is a variable of the $j$-th site's spin.
An illustration of the system is shown in Fig.~\ref{ising_instance}.
$\{ J_{jk} \}$ and $\{ h_j \}$ are random coefficients following the standard normal distribution.
It is difficult to calculate $\mu(\bm{x})$ using MCMC for this model, especially when the temperature $T$ is low.
Furthermore, no effective proposal distribution has been found for such a model, including the cluster update \cite{Houdayer_2001,Zhu_2015}.
The purpose of our numerical experiments is to demonstrate the advantage of our method using such a complex model that follows $\mu(\bm{x})$.

\subsection{Average convergence speed of optimized circuits}
\label{ave_conv_speed}
First of all, to investigate the performance of QAOA-MC, we analyze the absolute spectral gap $\delta$ of the Boltzmann distribution $\mu(\bm{x})$ for various model instances.
The temperature of the Boltzmann distribution is set to $T=0.1$.
We prepare 500 random spin glass instances by randomly choosing $\{ J_{jk} \}$ and $\{ h_j \}$ from a standard normal distribution and calculate $\delta$ for each $\mu(\bm{x})$.
The average convergence speed $\braket{\delta}$ for a model size of $n$ is obtained from these 500 $\delta$'s.
This is done for each $3 \leq n \leq 10$ to investigate the relationship between $n$ and $\braket{\delta}$.
We use the circuit in Fig.~\ref{quantum_circuit} with $p=5$ and set the hyperparameter $\theta_{\rm{max}}$ to 0.3 (see Appendix~\ref{appendix}).
In this numerical experiment, we compare our proposal to three proposal distributions: local update, uniform update, and ``random circuit''.
This ``random circuit'' corresponds to a distribution defined by our circuit (Eq.~\eqref{eq:u_proposed}) with a randomly chosen parameter $\theta \in [0, 2\pi]$ to verify the improvement of convergence speed through optimization.
This numerical experiment is simulated entirely on a classical computer using Python.
Qulacs \cite{suzuki2021qulacs} is utilized to simulate the quantum circuit.
The optimization method used is L-BFGS-B \cite{byrd1995limited}, which is implemented by Scipy \cite{virtanen2020scipy}.
$\mathrm{AR}$ is calculated exactly by Eq.~\eqref{eq:AR}.

\begin{figure}
    \centering
    \includegraphics[width=\linewidth]{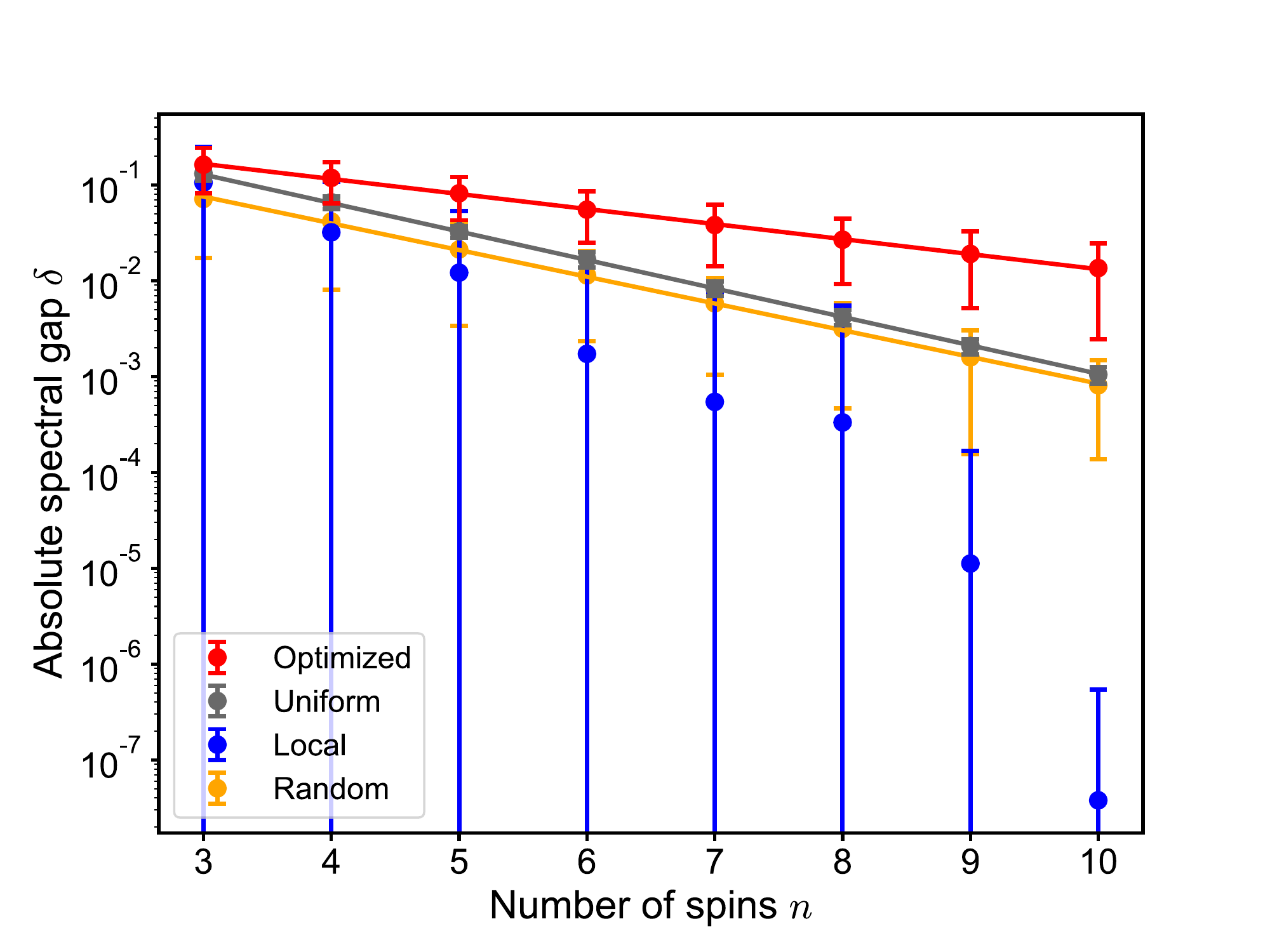}
    \caption{Relationship between model size $n$ and average convergence rate $\braket{\delta}$. ``Optimized'' is QAOA-MC, ``Uniform'' is the uniform update, ``Local'' is the local update, and ``Random'' uses the randomly chosen parameter $\theta \in [0, 2\pi]$ in our circuit.}
    \label{ave_gap_vs_qubit}
\end{figure}
Figure~\ref{ave_gap_vs_qubit} shows the relationship between $n$ and $\braket{\delta}$ obtained from the numerical experiment.
The points represent $\braket{\delta}$ computed using 500 random instances at each value of $n$.
The error bars represent the standard deviations computed over 500 $\delta$'s.
Although $\braket{\delta}$ decreases as $n$ increases for all methods, QAOA-MC shows a slower rate of decrease compared to others and is superior in terms of $\braket{\delta}$.
We fit $\braket{\delta}$ by $2^{-kn}$ with a parameter $k$ and show the result as the straight lines in Fig.~\ref{ave_gap_vs_qubit}.
The approximation curves fit the data well except for the local update.
The fitting is calculated using the least squares method.
Table~\ref{table_fitting} shows the scaling factor $k$ for these curves.
Uncertainties in Table~\ref{table_fitting} are obtained from the covariance matrices obtained in the fitting process.
\begin{table}[t]
    \caption{The value of the scaling factor $k$ obtained by fitting $\braket{\delta}$ with $2^{-kn}$.}
    \label{table_fitting}
    \centering
    \hbox to\hsize{\hfil
    \begin{tabular}{ccc}
      \hline
      Proposal  & $k$  &  Ratio to Uniform \\
      \hline \hline
      Optimized  & $0.521(6)$ & $1.89(2)$ \\
      Random  & $0.926(9)$ & $1.06(1)$ \\
      Uniform  & $0.9869(11)$ & $1$ \\
      \hline
    \end{tabular}
    \hfil}
\end{table}
QAOA-MC (``optimized'') has a scaling factor $k$ approximately 1/1.89 times that of the uniform update, which represents an approximately quadratic acceleration with respect to $\braket{\delta}$.
On the other hand, the results of the ``random circuit'' are almost identical to those of the uniform update, suggesting that this acceleration is due to the optimization of the circuit.

\begin{figure}
    \centering
    \includegraphics[width=\linewidth]{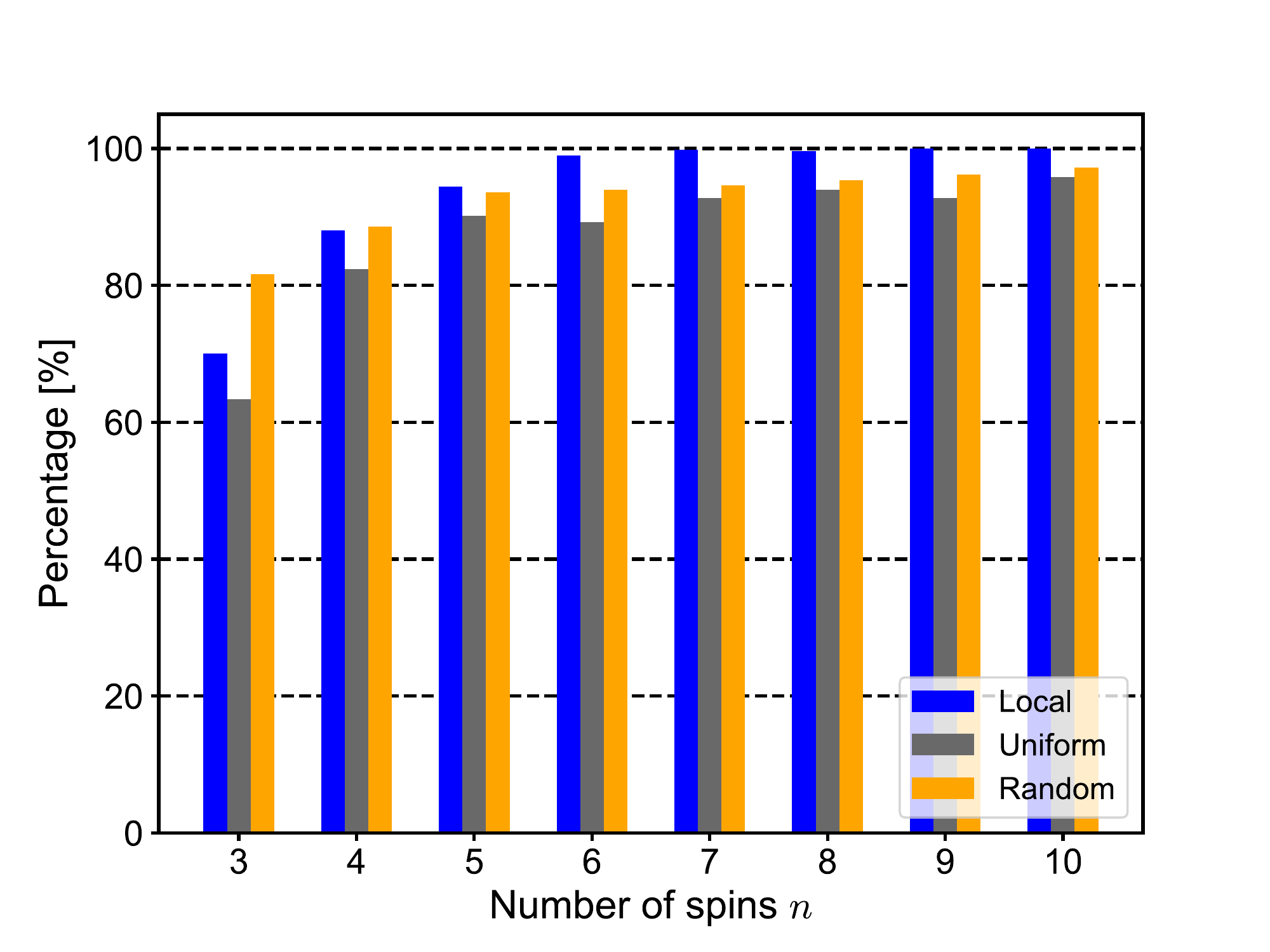}
    \caption{Percentage of instances in which our algorithm outperforms others in terms of convergence speed, for 500 random examples at each $n$.}
    \label{ave_result_2}
\end{figure}
Although QAOA-MC optimizes a parameter based on the observation that a local minimum of AR often gives a local maximum of $\delta$, it does not always hold true for all instances.
To see the effect of this imperfect assumption, we next show the percentage of instances for which QAOA-MC surpasses $\delta$ of other methods for each size $n$ in Fig.~\ref{ave_result_2}.
The percentage of instances in which our algorithm is dominant increases with increasing $n$.
For $n\geq7$, QAOA-MC outperforms the others in more than $90\%$ of the 500 instances, making it the best-performing proposal in this experiment.
This result indicates that our optimization method, which searches for a local minimum of AR for small values of $\theta$, works for many instances.

\subsection{Optimization with MCMC estimator of AR}
We now examine the impact of MCMC estimation of AR on the performance of QAOA-MC.
Since QAOA-MC must use MCMC estimate for obtaining AR in practice, the objective function contains statistical errors that could adversely affect the convergence performance.
We analyze the relationship between the number of samples used in AR estimation and the performance.
The numerical experiments performed here are under the same setup as discussed in Sec.~\ref{ave_conv_speed} unless otherwise stated.
AR is estimated from $M$ samples obtained through MCMC via Eq.~\eqref{eq:MCMC-AR}.
We set $M$ to $8$, $32$, $128$, and $\infty$ (where AR is calculated directly from the target distribution via Eq.~\eqref{eq:AR}) and optimize $\theta$.
Then, using the optimized $\theta$, we calculate the absolute spectral gap $\delta$ for the same instances used in Sec.~\ref{ave_conv_speed}.
When using MCMC estimators, the L-BFGS-B method cannot be used as the optimization method because the objective function contains statistical errors.
Here, we employ the bisection method for optimization, taking advantage of the fact that we only have a single parameter $\theta$.
In the numerical experiments, we used Brent's method \cite{brent2013algorithms}, which is implemented by Scipy.
The circuit and the hyperparameter settings are the same as in Sec.~\ref{ave_conv_speed}.

\begin{figure}
    \centering
    \includegraphics[width=\linewidth]{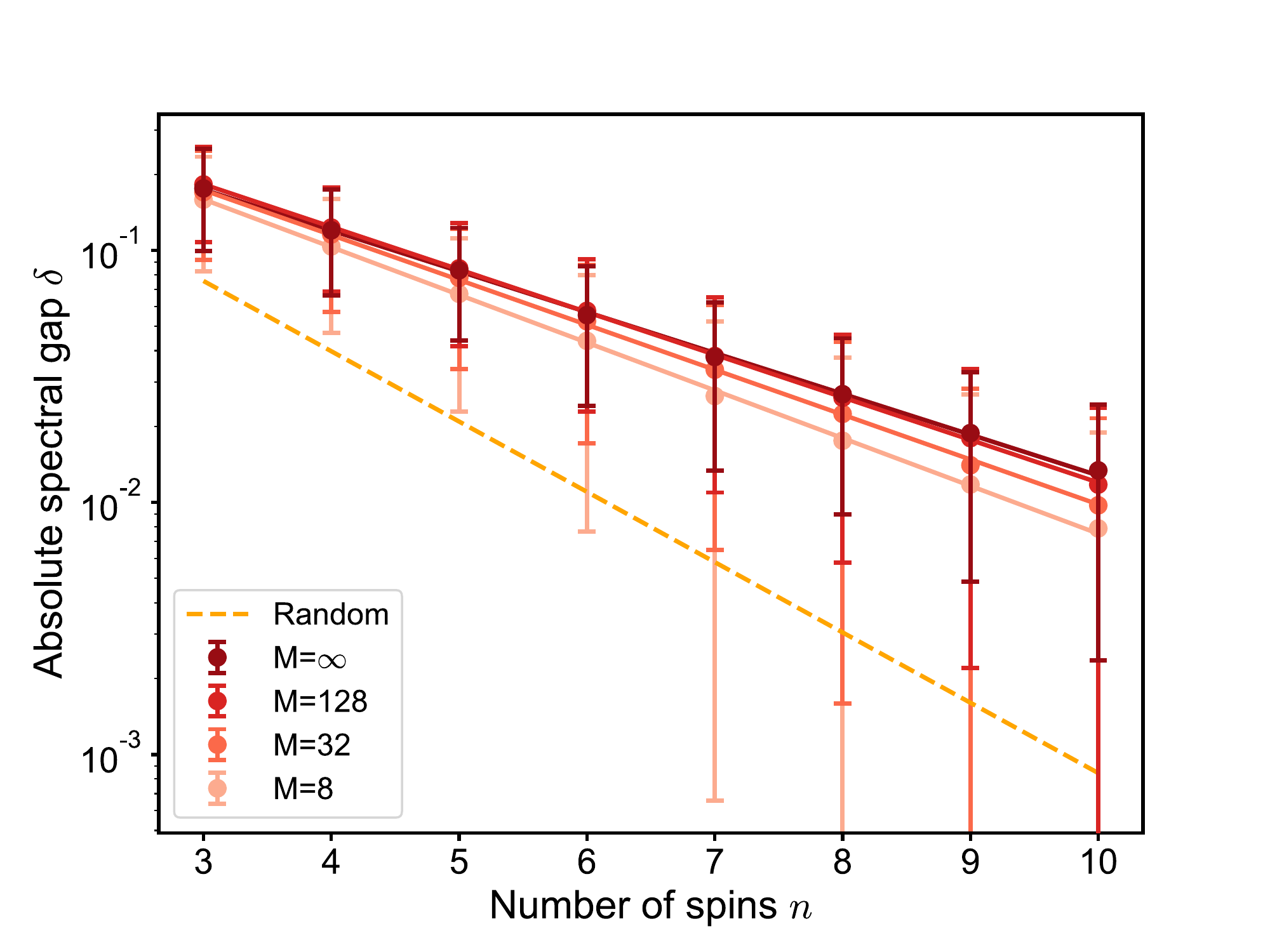}
    \caption{Relationship between model size $n$ and average convergence rate $\braket{\delta}$ when $M$ varies. The dotted line is the result of ``random'' in Fig. \ref{ave_gap_vs_qubit}.}
    \label{ave_gap_vs_M}
\end{figure}
\begin{table}[t]
    \caption{The value of the scaling factor $k$ of the approximate curve $\braket{\delta} \approx 2^{-kn}$ in our method when $M$ varies.}
    \label{table_fitting_2}
    \centering
    \hbox to\hsize{\hfil
    \begin{tabular}{ccc}
      \hline
      $M$  & $k$  &  Ratio to Uniform \\
      \hline \hline
      $\infty$  & $0.538(6)$ & $1.834(19)$ \\
      128  & $0.5615(26)$ & $1.758(9)$ \\
      32  & $0.592(6)$ & $1.668(17)$ \\
      8  & $0.629(6)$ & $1.568(15)$ \\
      \hline
    \end{tabular}
    \hfil}
\end{table}
Figure~\ref{ave_gap_vs_M} displays the relationship between $M$ and resulting $\braket{\delta}$.
Table~\ref{table_fitting_2} shows the scaling factor $k$ for the approximate curves obtained by the same fitting as Fig.~\ref{ave_gap_vs_qubit}.
As $M$ becomes smaller, the standard deviation of $\braket{\delta}$ increases and the scaling factor $k$ deteriorates at the same time.
This is because decreasing $M$ results in a less accurate AR estimate. 
It then leads to poor optimization, making the result towards that of ``random''.
On the other hand, if $M$ is large enough, the Brent method can be used to achieve a performance that is nearly the same as that attained by the L-BFGS-B method.
Note that the size of a sufficient $M$ in QAOA-MC is much less than the number of measurements used for a single evaluation of an expected value in VQAs.
Additionally, after the parameters have been determined by the optimization, only a single-shot measurement from the optimized circuit is needed for a single step of MCMC.

Finally, we test QAOA-MC by estimating a physical quantity.
The physical quantity to be estimated is the average magnetization $\langle m\rangle$ of the Ising model.
If the model follows the Boltzmann distribution $\mu(\bm{x})$, the average magnetization is defined as
\begin{align}
    \braket{m} = \sum_{j} \mu({\bm{x}}_j) m({\bm{x}}_j),
    \label{mag_ave}
\end{align}
where $m(\bm{x}) = \frac{1}{n} \sum_{j=1}^n x_j$ represents the magnetization of a state $\bm{x}$.
In this numerical experiment, we estimate the average magnetization of an $n=15$ spin glass instance with respect to the Boltzmann distribution at $T=1.0$.
The number of samples for AR estimation is set to $M=1000$.
The main MCMC simulation with optimized $\theta$ is conducted ten times, each starting with a random spin configuration and using 1000 steps.

\begin{figure}
    \centering
    \includegraphics[width=\linewidth]{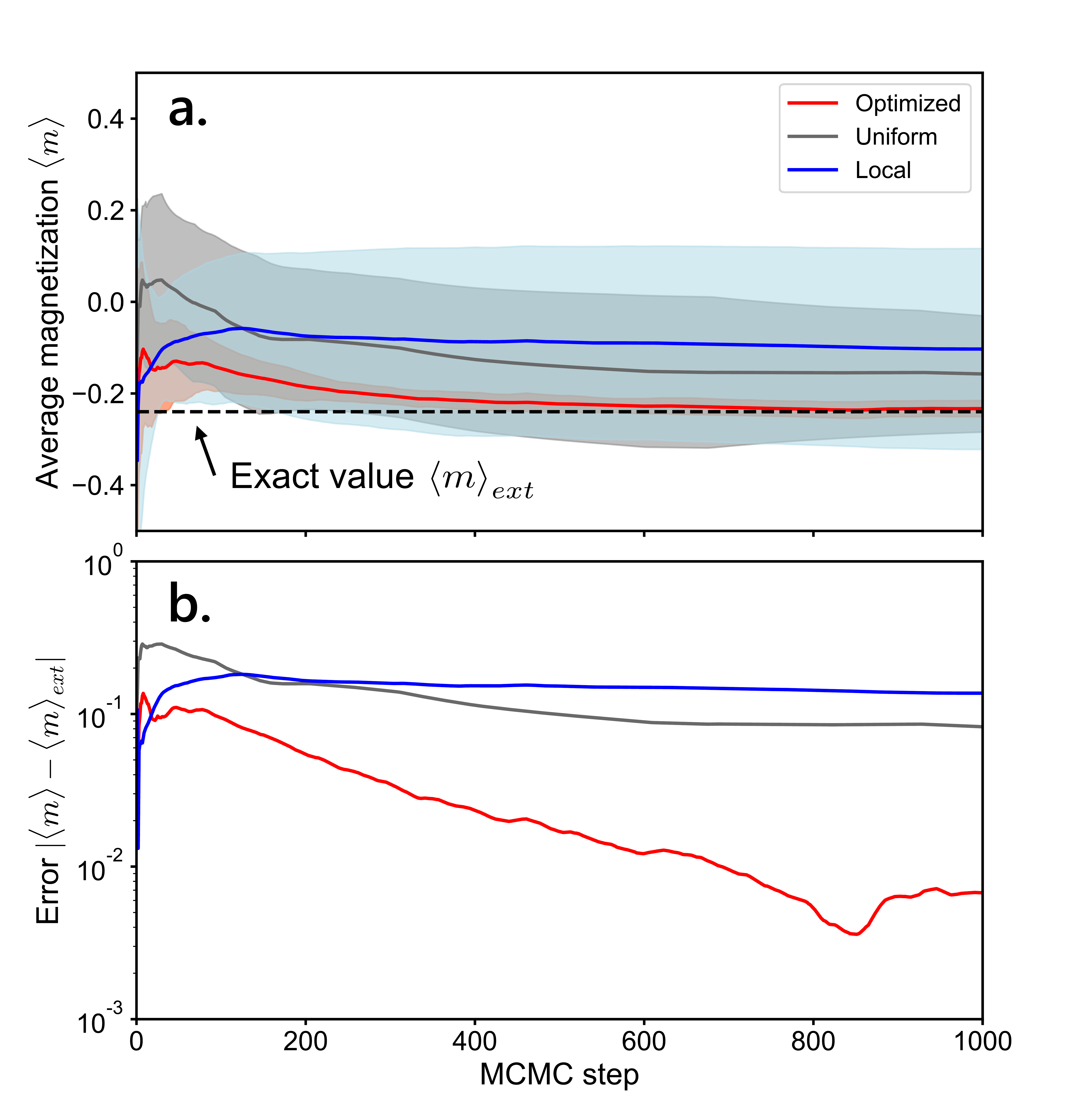}
    \caption{Estimation of the average magnetization $\braket{m}$ using MCMC. \textbf{a.} The relationship between the number of MCMC steps and the estimated value. \textbf{b.} The difference between the mean of the estimated value and the exact value.}
    \label{mcmc_result}
\end{figure}
Figure~\ref{mcmc_result} displays the MCMC estimation result of $\braket{m}$.
The solid lines represent the average evolution of $\braket{m}$ calculated from ten independent Markov chains at each step.
The bands indicate their standard deviations.
The dotted line represents the exact value of $\braket{m}$, which is calculated from the target distribution $\mu(\bm{x})$.
It can be seen that the average value of $\braket{m}$ converges faster to the dotted line in QAOA-MC compared to the others. Furthermore, the standard deviation of $\braket{m}$ in our algorithm is small and stable.

\section{Conclusion}
\label{Conclusion}
In this paper, we proposed QAOA-MC which uses samples from quantum circuits in the form of QAOA as the proposal for MCMC. 
Quantum computation is used only for proposing transitions and the other parts of the algorithm are executed on classical computers, which makes the algorithm feasible on current NISQ devices.
We introduced the use of a QAOA-type circuit to realize the algorithm with shallow circuits.
Furthermore, we showed that the convergence speed of MCMC can be improved by finding a local minimum of the AR.
As shown in numerical experiments, QAOA-MC confirmed an approximately quadratic speed-up in the absolute spectral gap for the Boltzmann distribution in spin glass, when compared to the uniform distribution.

Some future directions are in order.
First, our circuit in Fig.~\ref{quantum_circuit} originally has multiple parameters which could be further tuned to achieve a better proposal distribution.
However, the optimization of multiple parameters using AR did not work in our trials.
Building more advanced optimization methods remains a possible future work.
Additionally, the results of this study assume an ideal quantum computer without noise.
It is unclear whether the acceleration can be achieved on real NISQ devices.
If this advantage can be maintained despite the noise, our method has the potential to become a practical algorithm for current quantum computers.
Our algorithm leaves much room for improvement; in any case, it represents a new step toward implementing MCMC with quantum computers and facilitates the use of current NISQ devices.

\begin{acknowledgments}
This work is supported by MEXT Quantum Leap Flagship Program (MEXT Q-LEAP) Grant No. JPMXS0118067394 and JPMXS0120319794, and
JST COI-NEXT Grant No. JPMJPF2014.
KM is supported by  JST
PRESTO Grant No. JPMJPR2019.
\end{acknowledgments}

\appendix
\section{The choice of hyperparameter $\theta_{\rm{max}}$}
\label{appendix}
In this section, we analyze the distribution of parameters $\theta^*$ that give the locally minimal AR for various instances to determine the hyperparameter $\theta_{\rm{max}}$.
We define $\theta^*$ as $\theta$ that achieves the local minimum of AR, satisfies $\theta^* > 0$, and is closest to 0.
In our work, we use the average of $\theta^*$ in various instances as a hyperparameter $\theta_{\rm{max}}$.

\subsection{Analyzing $\theta^*$ vs. instance}
\label{appendix_a1}
\begin{figure}
    \centering
    \includegraphics[width=0.9\linewidth]{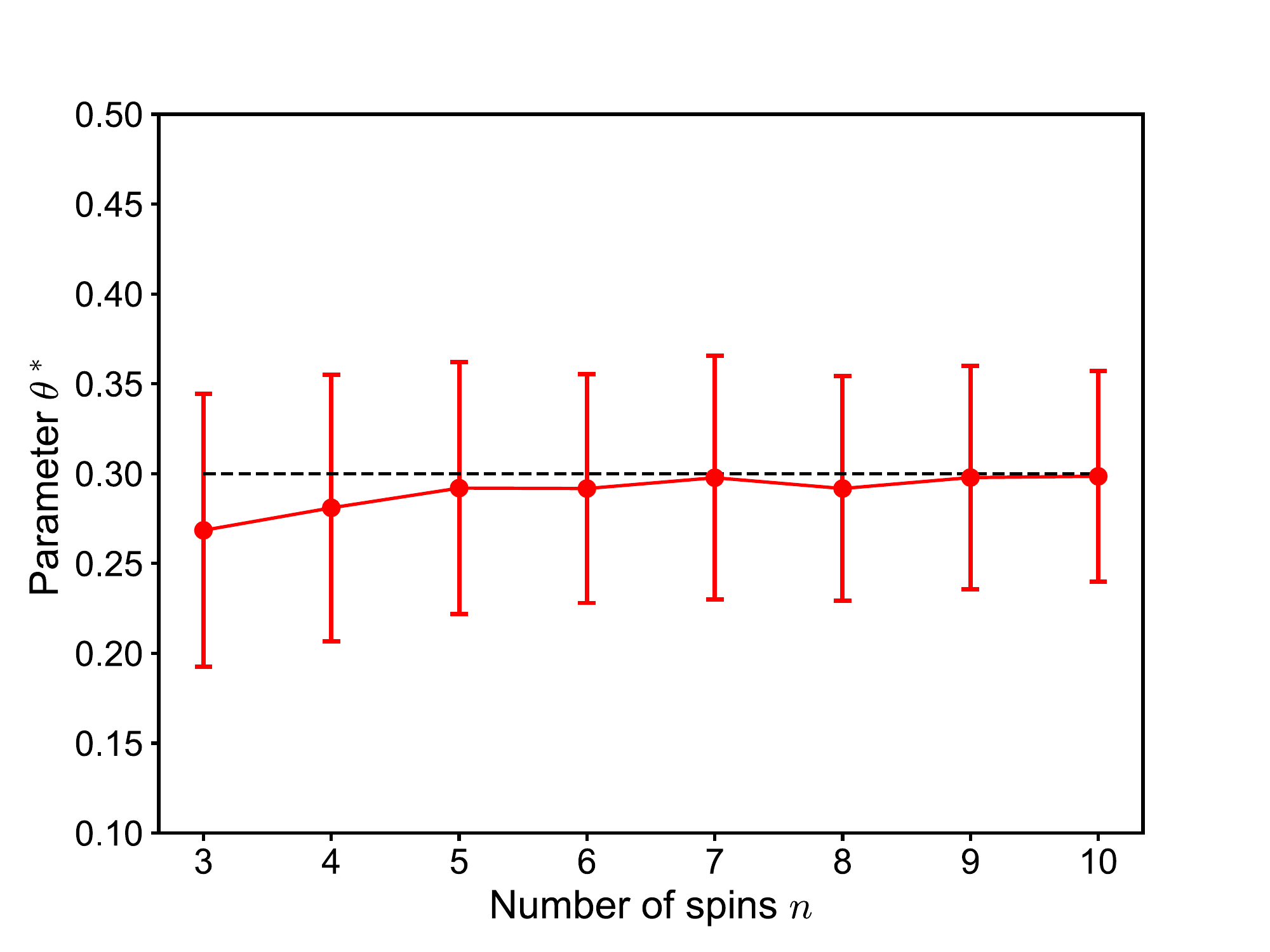}
    \caption{Relationship between $\theta^*$ and $n$ calculated by 500 random instances.}
    \label{appendix_1}
\end{figure}

$\theta^*$ varies with specific model instances.
To examine the distribution of $\theta^*$ among various instances, we generate 500 instances (Eq.~\eqref{eq:ising}) with random $\{ J_{jk} \}$ and $\{ h_j \}$ for each $3 \leq n \leq 10$ and determine $\theta^*$ for each of them.
In this numerical experiment, we use the circuit of Fig.~\ref{quantum_circuit} with $p = 5$.
The target distribution is the Boltzmann distribution with $T = 0.1$ (Eq.~\eqref{eq:Boltzmann_dist}).
The result is shown in Fig.~\ref{appendix_1}.
The dots denote the average, $\braket{\theta^*}$, of the 500 instances for each $n$, while the bands represent the corresponding standard deviations.
It can be seen that the average of $\braket{\theta^*}$ remains around 0.3 (indicated by a dotted line in Fig. \ref{appendix_1}) irrespective of $n$.
We, therefore, set $\theta_{\rm{max}} = 0.3$ in the numerical experiments presented in the main text.

\subsection{Analyzing $\theta^*$ vs. $p$}
\begin{figure}
    \centering
    \includegraphics[width=\linewidth]{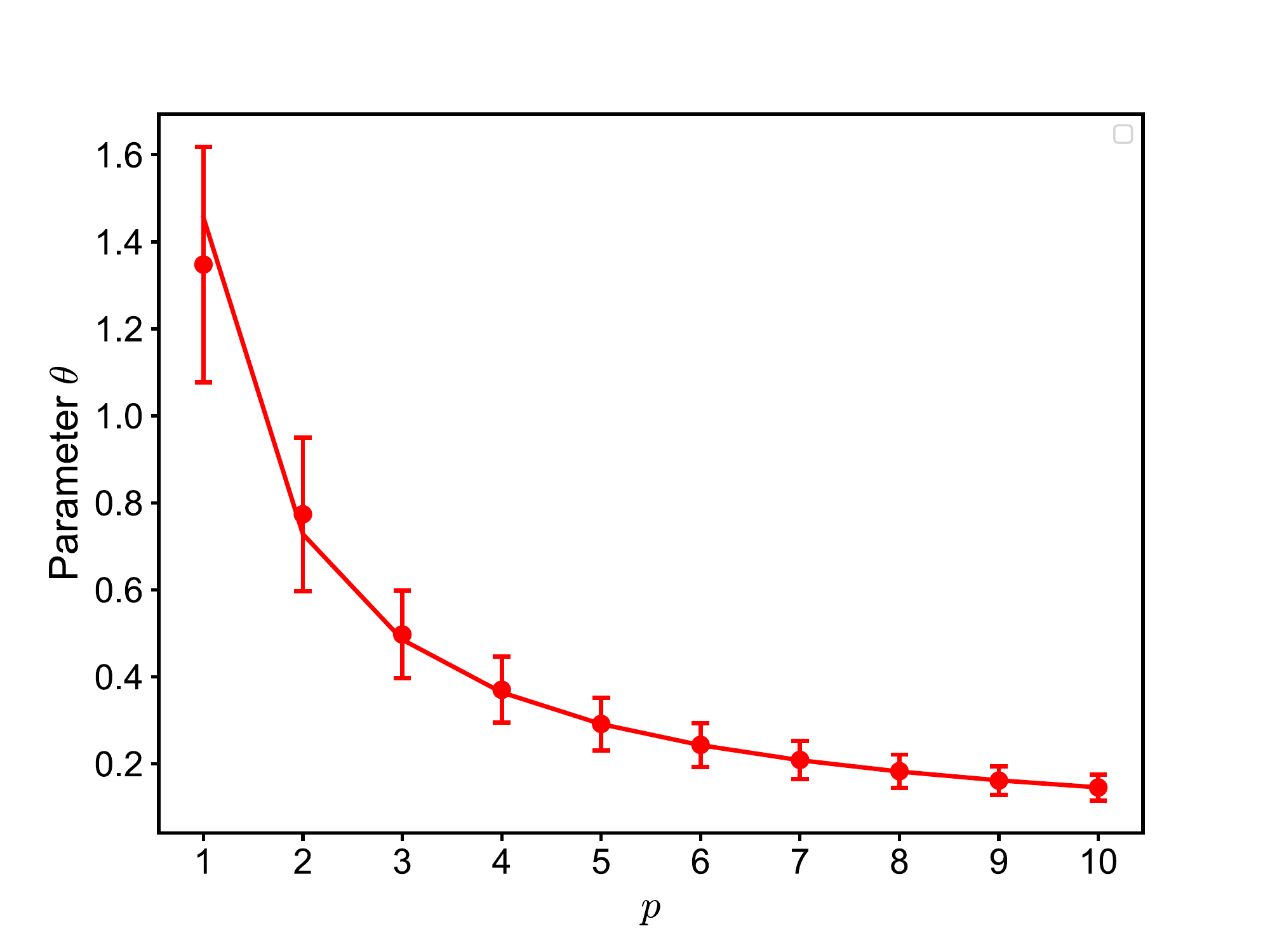}
    \caption{Relationship between $\theta^*$ and $p$ for each of the 50 random instances.}
    \label{appendix_3}
\end{figure}

We investigate the relationship between $p$ and $\theta^*$.
From Appendix~\ref{appendix_a1},  we see that $\theta_{\rm max}$ can be set at the same value, regardless of model instances.
However, this is not the case when $p$ varies.
We prepare 50 random instances for the spin glass (Eq.~\eqref{eq:ising}) for $n = 5$, vary the $p$ from $1$ to $10$, and calculate $\theta^*$.
The results are shown in Fig.~\ref{appendix_3}.
The dots denote the average $\theta^*$ of the 50 instances for each $p$, while the bands represent the corresponding standard deviations.
$\braket{\theta^*}$ is approximately proportional to $1/p$;
the curved line in Fig.~\ref{appendix_3} represents $a/p$ with $a=1.45558(25)$ which is obtained by using the least squares method.
When varying $p$, it seems appropriate to select $\theta_{\rm{max}}$ according to the fitting curve displayed in Fig.~\ref{appendix_3}.

\bibliography{cite}

\end{document}